\documentclass[showpacs,twocolumn,prd,nofootinbib]{revtex4-1}
 \usepackage{color,graphicx,epsfig}
 \usepackage{amsmath,amssymb,dsfont,epsfig,graphicx,xcolor}
 \usepackage{ifpdf}
 \usepackage{amsmath}
 \usepackage{bm}
 \usepackage{color}
 \usepackage[english]{babel}
 \usepackage{graphicx}%
 \usepackage{amsfonts}%
 \usepackage{amssymb}
 \usepackage{braket}
 \usepackage{hyperref}
 \usepackage{epstopdf}
 \usepackage{slashed}
\usepackage{float}
 \graphicspath{{Plots/}}
 \usepackage{commath}
 
\begin{document}

\def\LjubljanaFMF{Faculty of Mathematics and Physics, University of Ljubljana,
 Jadranska 19, 1000 Ljubljana, Slovenia }
\def\LjubljanaIJS{Jo\v zef Stefan Institute, Jamova 39, 1000 Ljubljana, Slovenia}
\def\Orsay{Universit\'e Paris-Saclay, CNRS/IN2P3, IJCLab, 91405 Orsay, France}

\title{  Decay dynamics of $N\to \ell \pi $  and $N\to \ell \gamma$ }

\author{Svjetlana Fajfer}
\email[Electronic address:]{svjetlana.fajfer@ijs.si} 
\affiliation{\LjubljanaIJS}
\affiliation{\LjubljanaFMF} 

\author{Mitja Sadl}
\email[Electronic address:]{mitja.sadl@fmf.uni-lj.si} 
\affiliation{\LjubljanaFMF} 

\date{\today}
\begin{abstract}
SuperKamiokande experiment tightly bounds the lifetimes of the baryon number violating proton decays.  
The decay widths for the nucleon to an antilepton and a photon are not so well bounded experimentally as $\ell \pi$ modes. Using an effective Lagrangian approach, we relate the decay widths of $N\to \ell \gamma $ to the decay widths of $N\to \ell \pi$. 
Our result points out  factor $10^{3}$ suppression of the decay widths $\Gamma (p   \to \ell^{+} \gamma)$ and $\Gamma (n \to \bar \nu  \gamma)$ compared to the decay widths $\Gamma (p  \to \ell^{+}  \pi^{0})$ and $\Gamma (n \to \bar \nu \pi^{0})$, respectively. This result is independent of the model of new physics.   
Then, we investigate the dynamics of   $N\to \ell \pi $ and  $N\to \ell \gamma$ amplitudes at the tree and loop level in which scalar leptoquarks are mediators of a new interaction.
At the tree level, leptoquarks probe the physics beyond the Standard Model at a scale of $10^{16}$ GeV, while at the loop level, such decay can occur at a scale of $10^{7}$ GeV.  \end{abstract}

\maketitle

\section{Introduction} 
In the Standard Model (SM), the baryon number is conserved due to an accidental symmetry,  contrary to many beyond SM (BSM) approaches that violate the baryon number. Conservation of the baryon number is experimentally tested in many decay modes, and it was found that the proton is stable up to $10^{34}$ years {\cite{ParticleDataGroup:2022pth}. The lifetime of the decay $p\to e^+ \pi^{0}$ is the most constrained among all decay modes \cite{FileviezPerez:2022ypk}. On the theory side, this  decay mode was 
investigated in detail within a variety of approaches (see e.g.  \cite{Weinberg:1980bf,Nath:2006ut,Heeck:2019kgr,Hambye:2017qix,Fonseca:2018ehk,FileviezPerez:2019ssf,Helo:2019yqp}).
The authors of Refs. \cite{Weinberg:1980bf,Nath:2006ut,Heeck:2019kgr,Hambye:2017qix} used the effective Lagrangian approach without specifying a particular beyond SM (BSM) theory. The dimension-six operators can generate the nucleon decays to pseudoscalar meson and lepton, as in \cite{Weinberg:1980bf,Nath:2006ut,Heeck:2019kgr,Hambye:2017qix}. However, they can be generated by the higher dimension operators or the dimension-six operators resulting from the loop diagrams \cite{Dorsner:2012nq,Helo:2019yqp}. 

 The same operators inducing nucleon to antilepton and pion decays are responsible for the radiative decay mode. 
The first studies of proton radiative decays were done a long time ago \cite{Eeg:1981be,Silverman:1980ha}. Recently the authors of  Ref. \cite{Bansal:2022xbg}  reconsidered $p\to \ell^+ \gamma$ using light-cone QCD sum rules.
The lattice QCD greatly improved the calculation of the matrix elements for the nucleon to pion transitions \cite{Aoki:2017puj}. They also provide the matrix elements of the relevant dimension-six operators between nucleon and vacuum state. The dimension six operators with the flavour structure $(u u d \ell)$ and  $(u  d d \ell)$ are present in both transitions. Namely,  we can use the lattice  QCD knowledge of the form factors present in $N\to \pi$ transition and the annihilation amplitude  $N\to \ell$, and therefore reconsider radiative decays $N\to \ell \gamma$  and relate their decay widths to the amplitudes for $N\to \ell \pi$.

Moreover, the authors of Refs. \cite{Fornal:2020bzzV2,Fornal:2020gtoV2,Fornal:2018eol} suggested explaining the difference in the neutron lifetime, measured in the beam and bottle experiment, known as neutron decay anomaly.  
They analysed the neutron transition to a dark fermion by mixing the neutron and the dark fermion. 
We extend this approach to the transition of a proton to a charged lepton or a neutron to an anti-neutrino. In Table \ref{tab-exp}, we list experimental bounds on the transitions we consider in this paper. 
\begin{table}[h]
\centering
\begin{tabular}{|c|c|}
\hline
Decay mode&  $\Gamma^{-1}/10^{30} yr$ \\
\hline
$p\to e^+ \pi^0$ & 16000 \cite{Miura:2016krn}\\
$p\to \mu^+ \pi^0$ & 7700 \cite{Miura:2016krn}\\
$n\to \nu\, \pi^0 $& 1100 \cite{Abe:2013lua}\\
$p\to e^+ \gamma$ & 670 \cite{Sussman:2018ylo}\\
$p\to \mu^+ \gamma$ & 478 \cite{Sussman:2018ylo}\\
$n\to \nu\, \gamma $& 550 \cite{Takhistov:2015fao}\\
\hline 
\end{tabular}
\caption{\label{tab-exp} Experimental lower bounds on nucleon's partial mean lifetime.}
\end{table}
First, we consider the transition of the nucleon to a lepton and a photon,  in a general framework, based on the transition of the nucleon to a lepton, due to the dimension-6 Lagrangian, which generates baryon number violation. 
Then we consider specific examples of the baryon number-violating models. The simplest extensions of the SM are those with scalar leptoquarks. Scalar leptoquarks can be easily incorporated in ultraviolet complete frameworks (see, e.g. \cite{Dorsner:2016wpm}). On the other hand, vector leptoquarks can be treated as gauge bosons, and a complete theoretical framework should be known if the loop diagrams are considered.
In Section \ref{sec:2}, we present the theoretical framework of our calculations. Section \ref{sec:3} contains the leptoquark contributions analysis to $N\to \ell \pi$ and $N\to \ell \gamma$  decay amplitudes at tree-level and loop-level transitions. Numerical results are presented in Section \ref{sec:4}, while Section \ref{sec:5} contains a summary of our results.

\section{Framework}
\label{sec:2}

First, we consider the effective Lagrangian approach as a framework to describe the nucleon decay $N\to \ell \pi$ and $N\to \ell \gamma$ and then collect the inputs from the Lattice QCD, which enables us to relate the decay widths of the radiative decays $N\to \ell \gamma$ to the decay widths of $N\to  \ell \pi$. 

\subsection{Effective Lagrangian  for  $\Delta B=1$ transitions} 

The operators describing $\Delta B=1$  transitions appear at mass dimension six. Making their flavour structure explicit, these operators  for the first generation of quarks and leptons can be written as
\begin{align}
\mathcal{L}_{d=6} &=  \frac{C_1}{\Lambda^2}
\epsilon^{\alpha\beta\gamma} \epsilon_{il}\epsilon_{jk} (\overline{Q}^C_{i ,\alpha} Q_{j,\beta})(\overline{Q}^C_{k,\gamma}  L_{l}) \nonumber\\
&\quad +\frac{C_2}{\Lambda^2}\epsilon^{\alpha\beta\gamma} (\overline{Q}^C_{i,\alpha}\epsilon_{ij} Q_{j,\beta})(\overline{u}^C_{\gamma} \ell) \nonumber\\
&\quad + \frac{C_3}{\Lambda^2} \epsilon^{\alpha\beta\gamma}(\overline{d}^C_{\alpha} u_{\beta})(\overline{Q}^C_{i,\gamma}\epsilon_{ij} L_{j}) \nonumber\\
&\quad +\frac{C_4}{\Lambda^2} \epsilon^{\alpha\beta\gamma} (\overline{d}^C_{\alpha} u_{\beta})(\overline{u}^C_{\gamma} \ell)+\text{h.c.} \,,
\label{L-6}
\end{align}
where  $\alpha,\beta,\gamma$ denote the colour, $i,j,k,l$ the $\mathrm{SU}(2)_L$ indices, ~\cite{Weinberg:1979sa,Wilczek:1979hc,Weinberg:1980bf,Weldon:1980gi,Abbott:1980zj}.
The letters $u$, $d$, and $\ell$ denote the right-handed up-quark, down-quark, and lepton fields, while $Q$ and $L$ are the left-handed quark and lepton doublets, respectively.                                                                                                                                                          
 The $\Lambda$ parameter denotes a mass scale of mediators while  $C_j$ couplings have zero dimension.  
 Such a dimension-six operator might arise from the tree-level or loop-level interaction. The  effective Lagrangian in Eq. (\ref{L-6}) describes both transitions $N\to \ell \gamma$ and $N\to \ell \pi$. 

\subsection{$N\to \ell \pi$ decays} 

We briefly summarise basic results on  $p\to \ell^+ \pi^0 $ ($n \to \bar \nu \pi^{0}$) decay amplitudes coming from Lagrangian in Eq. (\ref{L-6}) and their decay widths.  To calculate the transition amplitude, one must first determine the matrix elements of the operators between the nucleon and pion. These inputs are delivered by the lattice QCD.  The standard lattice QCD parametrisation for the nucleon to pion transition is \cite{Aoki:2017puj,Aoki:2006ib} 
\begin{equation}
\left\langle P \left|\mathcal{O}^{\Gamma\Gamma'}\right|N\right\rangle=\left[W_0^{\Gamma\Gamma'}(q^2)-\frac{i\slashed{q}}{m_N}W_1^{\Gamma\Gamma'}(q^2)\right]P_{\Gamma'}u_N\>,
\label{E87a}
\end{equation}
where 
\begin{equation}
\mathcal{O}^{\Gamma\Gamma'}=\left(\overline{q}^{\textrm{C}}P_{\Gamma}q\right)P_{\Gamma'}q\qquad \textrm{and}\qquad \Gamma, \Gamma'=R,L.
\label{E87b}
\end{equation}
In eq. (\ref{E87a})  $u_{N}$ is the nucleon spinor, $N=p,\, n$. 
The operators $\mathcal{O}^{\Gamma\Gamma'}$ represent one of the operators in eq. \ref{L-6}, and $\Gamma, \, \Gamma'$ denotes the handedness of the operators. 
In the proton decay amplitude,  the form factor $W_1^{\Gamma\Gamma'}(q^2)$ is small compared to $W_0^{\Gamma\Gamma'}(q^2)$ and thus neglected \cite{Aoki:2017puj}. 
The decay width $p\to \ell^+ \pi^0 $ comes from the first operator in (\ref{L-6}).\begin{eqnarray}
&&\Gamma(p\to \ell^+ \pi^0) = \frac{1}{32 \pi} \left| \frac{ C_1^p  }{\Lambda^2} \right|^2\nonumber\\
&& \times \left( W_0^{\textrm{LL}}(0)  \right)^2 
 \left( m_{p}^{2} -m_{\pi}^{2} +m_{\ell}^{2 } \right) \frac{\lambda^{1/2}(m_{p}^{2}, m_{\ell}^{2}, m_{\pi}^{2})}{m_{p}^{3} }\nonumber\\
 \label{pepil}
\end{eqnarray}
As usual $\lambda(a,b,c) = a^2+b^2+c^2 -2(a\,b +a\,c +c\,b) $ is the K\" all\' en function. 
The appropriate replacements should be done for the chiralities different from $\textrm{LL}$.

\subsection{Radiative decays $p \to \ell^{+ }  \gamma$  and $n \to \bar \nu \gamma$  } 

To approach nucleon decays to a lepton and photon, we extend the procedure suggested in Ref. \cite{Fornal:2018eol} for the radiative decay amplitude calculation and write a general effective Lagrangian for the nucleon radiative transitions. 
First, we write down the effective interaction of proton and $\gamma$
\begin{eqnarray}
&&{\cal L}_{p\gamma p} = e \bar p \left(  \slashed{A}   + \frac{a_p }{4\, m_p} \sigma^{\alpha \beta} F_{\alpha \beta} \right) p, \label{Int-p}
\end{eqnarray}
where $a_p$ is related to the proton's anomalous magnetic moment  $a_p= 1.793$. 
For the neutron it is
 \begin{eqnarray}
&&{\cal L}_{n\gamma n} = e \bar  n\left(  \frac{a_n }{4\, m_n} \sigma^{\alpha \beta} F_{\alpha \beta} \right) n, \label{Int-n}
\end{eqnarray}
with $a_n=- 1.913$. We allow the proton - positron  (neutron - antineutrino) mixing as suggested in Ref. \cite{Fornal:2018eol} 
\begin{equation}
{\cal L}_{\textrm{mix}}^{p\ell}= \varepsilon_{p} (\bar p \ell + \bar \ell p),\; \;
 {\cal L}_{\textrm{mix}}^{n\nu}= \varepsilon_{n} (\bar n \nu + \bar \nu n). 
\label{N-mix}
\end{equation} 
with $\varepsilon_{p}$, $\varepsilon_n$ being mixing parameters with a dimension of mass.
After the diagonalisation of the mass matrices, in the limit  $\varepsilon_p\ll ( m_p -m_\ell)$ or $\varepsilon_n\ll  m_n$  such interaction leads to the contributions \cite{Fornal:2018eol}
\begin{equation}
{\cal L}_{p\to \ell \gamma}^{\textrm{eff}}= -\frac{a_p e}{4\, m_p} \frac{\varepsilon_p}{m_p- m_\ell} \, \bar \ell \sigma^{\alpha \beta} F_{\alpha \beta} \, p+\textrm{h.c.} 
\label{eGp}
\end{equation} 

In the case of  $n \to \bar \nu \gamma$ decay,  $a_p$ should be replaced by $a_n$, as well as masses of proton and charged lepton by masses of neutron and zero-mass of the antineutrino
\begin{equation}
{\cal L}_{n\to \bar \nu  \gamma}^{\textrm{eff}}= -\frac{a_n  e \,\varepsilon_n}{4\, m_n^{2 }}  \bar \nu \sigma^{\alpha \beta} F_{\alpha \beta} \, n+\textrm{h.c.}
\label{eGn}
\end{equation} 

The numerical calculations of radiative decay amplitudes rely on the lattice QCD calculation  of the matrix elements \cite{JLQCD:1999dld,Aoki:2017puj,Aoki:2006ib}
\begin{eqnarray}
&&\langle 0| (ud)_R u_L|p\rangle = \alpha_p P_L u_p, \, \langle 0| (ud)_L u_R|p\rangle = -\alpha_p P_R u_p, \nonumber\\
&&\langle 0| (ud)_L u_L|p\rangle  =\beta_p P_L u_p, \, \langle 0| (ud)_R u_R|p\rangle  =-\beta_p P_R u_p,\nonumber\\
\label{ab}
\end{eqnarray}
where colour indices are not explicitly specified. From the Lagrangian (\ref{L-6}) we can calculate  $\varepsilon_{N}$. To specify which Wilson coefficient $C_{i}$ in eq. (\ref{L-6}) generates the mixing parameter, we use a more precise notation $\varepsilon_N \to \varepsilon_{i}^{N}$, where $\varepsilon_{i}^{N}= \alpha_{N} (\beta_{N})\,C_{i}/\Lambda^{2} $.\footnote{A change between $\alpha_N$ and $\beta_N$ will not change our numerical results since $|\alpha_N|=|\beta_N|$.} The chirality of the operators in (\ref{ab}) 
determines if $\alpha_{N}$ or $\beta_{N}$ are present. 
The advantage of this approach is that the radiative decay amplitude contains the matrix element of a nucleon annihilation to a lepton.  
The decay width for the radiative decays can be written as
\begin{eqnarray}
&&\Gamma(p\to \ell^+ \gamma)=\nonumber\\ 
&&=\frac{e^2 a^2_p }{32\pi}\left| \frac{ C_1^p  }{\Lambda^2} \right|^2\beta_p^2 
\frac{m_p}{(m_p-m_{\ell})^2}\left(1-\left(\frac{m_{\ell}}{m_p}\right)^2\right)^3. \nonumber\\
\label{N-L-gamma}
\end{eqnarray}
For $n\to \bar \nu \gamma$ appropriate replacement of masses, $m_{\ell} \to 0$ and $a_p\leftrightarrow a_n$ should be done in the above equation. $\beta_N$($\alpha_N$) is present if we have $C_1$ or $C_4$ ($C_2$ or $C_3$).
One can immediately express the decay width of $p\to \ell^{+} \gamma$  using the the relation for the the decay width of $p\to \ell^{+} \pi^{0}$
\begin{equation}
\Gamma(p\to \ell^+ \gamma)= C_{\gamma\ell} \, \Gamma(p\to \ell^+ \pi^0),
\label{G-L}
\end{equation}
with 
\begin{eqnarray}
&&C_{\gamma\ell} =e^2\, a^2_p \,\frac{ \beta_p^2}{W_0^{\textrm{LL}}(0)^2}F(m_p,m_{\ell},m_{\pi}),
\label{NL-comb}
\end{eqnarray}
and
\begin{eqnarray}
&&F(m_p,m_{\ell},m_{\pi} )=\nonumber\\
&&= \frac{m_p^4 \left(1-\left(\frac{m_{\ell}}{m_p}\right)^2\right)^3}{(m_p-m_{\ell})^2  \lambda^{1/2}(m_{p}^{2}, m_{\ell}^{2} m_{\pi}^{2} ) (m_p^{2}+m_{\ell}^{2} - m_{\pi}^{2})}.
\label{F-comb}
\end{eqnarray}
For the neutron decays, one should replace masses $m_{p}\to m_{n}$ and  $m_{\ell}\to 0$. Note that the proportionality factor $C_{\gamma\ell} $ contains the ratio of the two lattice results, the constant $\alpha_{N} (\beta_{N})$ and the form factor for the matrix element of the operators creating the nucleon - pion transition. 
The important result of our study is that by using the effective nucleon-lepton mixing approach  \cite{Fornal:2018eol},  one can relate the decay widths of the radiative mode to  
$\Gamma(N\to \ell \pi)$, independently on the model of new physics. 



\section{Scalar Leptoquarks in $N\to \ell \pi$ and $N \to \ell \gamma$ }
\label{sec:3}

Leptoquark, scalar or vector, mediates the interaction of a quark and a lepton. The fermion number $F = 3 B+L $ ($B$ is the quark baryon number, and $L$  stands for the lepton number) is useful in classifying leptoquarks. The leptoquark multiplets that couple to the quark $-$ lepton (antiquark $-$ lepton) pairs have the fermion number  $|F| = 2\, \, (F = 0)$ (for details see \cite{Dorsner:2016wpm}).  Leptoquarks having  $F = - 2$ can mediate proton decay at the tree level if diquark couplings are not forbidden. We consider nucleon decays induced by the scalar leptoquarks at the tree and loop levels. To include vector leptoquarks in the analysis, it is necessary to know the full ultraviolet theory containing them. That is behind the scope of our analysis.  The quantum numbers of leptoquarks regarding the SM colour, weak isospin and electromagnetic charges are specified in Table \ref{tab:LQs}. When the $\Delta B=1$ transition occurs at the tree-level, only 
$S_{1}^{1/3}$ and $S_{3}^{1/3}$ can generate amplitudes. The box diagram can be generated with $S_{3}^{4/3}= (\bar 3, 3, 4/3)$, $S_{3}^{-2/3}= (\bar 3, 3, -2/3)$  and $\tilde S_{1}= (\bar 3, 1, 4/3)$ accompanied by $W$ gauge bosons mediating interactions.  
The triple-leptoquark interactions generate a special case of loop diagrams, which can generate decays of nucleons to three leptons at tree level, as discussed in \cite{Dorsner:2022twk,Hambye:2017qix}. 
\begin{table}[tbp]
\centering
\begin{tabular}{|c|c|c|}
\hline
$(SU(3),SU(2),U(1))$ &  Symbol &  $\Delta B= 1$ nucleon decays \\
\hline 
$(\overline{\mathbf{3}},\mathbf{3},1/3)$ & $S_3^{1/3}$  &  tree \\
$(\overline{\mathbf{3}},\mathbf{3},1/3)$ & $S_3^{-2/3}$  & loop\\
$(\overline{\mathbf{3}},\mathbf{3},1/3)$ & $S_3^{4/3} $  & loop\\
$(\overline{\mathbf{3}},\mathbf{1},4/3)$ & $\tilde{S}_1$ & loop  \\
$(\overline{\mathbf{3}},\mathbf{1},1/3)$ &  $S_1$ &tree  \\
$(\overline{\mathbf{3}},\mathbf{1},-2/3)$ &  $\bar{S}_1$ & tree, to non-SM lepton\\
\hline
\end{tabular}
\caption{\label{tab:LQs} List of scalar leptoquarks.  The hyper-charge $Y$ normalisation is defined through $\hat{Q}=I_{3}+Y$, where $\hat{Q}$ is the electric charge operator, and $I_{3}$ is the third component of the weak isospin. The weak doublets $R_2$ and  $\tilde{R}_2$ can be part of the triple leptoquarks coupling.   } 
\end{table} 
The weak singlets $\bar S_{1}$ with the electric charge $-2/3$ can produce baryon decay amplitudes at tree level. However, the final state has quantum numbers of right-handed neutrinos, and their decays were considered in \cite{Fajfer:2020tqf,Fajfer:2021woc}. Two leptoquark weak doublets $R_2=(\mathbf{3},\mathbf{2},7/6)$ and $\tilde{R}_2=(\mathbf{3},\mathbf{2},1/6)$ with the fermion number $F = 0$ do not have diquark couplings and therefore do not lead to the nucleon decay at tree level. However, $R_2$ and  $\tilde{R}_2$ contribute to triple leptoquarks couplings. 
 In the following examples, we add a new superscript to $\varepsilon_{i}^{N}$ to denote which leptoquark is the mediator. By assuming $\Lambda \simeq M_{LQ}$,  $\varepsilon_{i}^{N} \to \varepsilon_{i}^{N,LQ} = \alpha_{N} (\beta_{N})\,C_{i}/M_{LQ}^{2} $.

\subsection{Scalar Leptoquarks $S_{1}$ and $S_{3}$ in nucleon decays at tree-level}

The $S_1$ leptoquark, as a weak singlet, can couple to left- and right-handed fermions.  Without specifying colour indices,  the interaction reads  \cite{Dorsner:2016wpm}
\begin{align}
\nonumber
& \mathcal{L}_{S_{1}} = -(y^{LL}_1 U)_{ij} \bar{d}_{L}^{C\,i} S_{1} \nu_{L}^{j}+(V^T y^{LL}_1)_{ij}\bar{u}_{L}^{C\,i} S_{1} e_{L}^{j}\\
\nonumber
&+y^{RR}_{1\,ij}\bar{u}_{R}^{C\,i} S_{1} e_{R}^{j}+y^{\overline{RR}}_{1\,ij}\bar{d}_{R}^{C\,i} S_{1} \nu_{R}^{j}\\
\nonumber
&+(V^T z^{LL}_1)_{ij}\bar{u}_{L}^{C\,i} S^*_{1} d_{L}^{j}-(z^{LL}_1 V^\dagger)_{ij}\bar{d}_{L}^{C\,i} S^*_{1} u_{L}^{j}\\
\nonumber
&+z^{RR}_{1\,ij}\bar{u}_{R}^{C\,i} S^*_{1} d_{R}^{j}+\textrm{h.c.}\,,\\
\label{eq:main_S_1_a}
\end{align}
where $U$ represents a Pontecorvo--Maki--Nakagawa--Sakata (PMNS) unitary mixing matrix and $V$ is a Cabibbo--Kobayashi--Maskawa (CKM) mixing matrix. 
The  $S_{3}$  scalar leptoquark is a weak triplet that only allows its coupling to left-handed fermions. This leptoquark has three components $S^{-2/3}_3$,
$S_{3}^{1/3}$ and $S^{4/3}_3$. The interacting Lagrangian with the matter fields is \cite{Dorsner:2016wpm}
\begin{align}
\nonumber
&\mathcal{L}_{S_{3}}= -(y^{LL}_{3}U)_{ij}\bar{d}_{L}^{C\,i} S^{1/3}_{3} \nu_{L}^{j}-\sqrt{2} y^{LL}_{3\,ij}\bar{d}_{L}^{C\,i} S^{4/3}_{3} e_{L}^{j}+\\
\nonumber
&+\sqrt{2} (V^Ty^{LL}_{3}U)_{ij}\bar{u}_{L}^{C\,i} S^{-2/3}_{3} \nu_{L}^{j}-(V^Ty^{LL}_{3})_{ij}\bar{u}_{L}^{C\,i} S^{1/3}_{3} e_{L}^{j}-\\
\nonumber
&-(z^{LL}_{3}V^\dagger)_{ij}\bar{d}_{L}^{C\,i} S^{1/3\,*}_{3} u_{L}^{j}-\sqrt{2}z^{LL}_{3\,ij}\bar{d}_{L}^{C\,i} S^{-2/3\,*}_{3} d_{L}^{j}+\\
\nonumber
&+\sqrt{2}(V^T z^{LL}_3 V^\dagger)_{ij}\bar{u}_{L}^{C\,i} S^{4/3\,*}_{3} u_{L}^{j}-(V^Tz^{LL}_3)_{ij}\bar{u}_{L}^{C\,i} S^{1/3\,*}_{3} d_{L}^{j}\\
&+\textrm{h.c.}
\label{eq:main_S_3_a}
\end{align}
The $N \to \ell \pi$ and $N\to \ell \gamma$ decay amplitudes can be generated at tree level by the exchanges of either $S_{1}^{1/3}$ or $S_{3}^{1/3}$, as shown in \cite{Dorsner:2012nq}.   
Note that  $S^{4/3}_3$ and $S^{-2/3}_3$ cannot contribute to the proton (neutron) decay at the tree level since the diquark couplings must be antisymmetric in flavour space. 

We determine the Wilson coefficients in the Lagrangian (\ref{L-6}). 
After integrating out the $S_1$  contribution, setting the scale $\Lambda$ to be equal $m_{S_1}$,  the corresponding Wilson coefficients are 
\begin{eqnarray}
&& C_1^{p,S_1}=  (V^T z^{LL}_1)_{11}  (V^T  y_1^{LL} )_{11} ,\nonumber\\
&&C_2^{p, S_{1}}=  (V^T z^{LL}_1)_{11}  (y_1^{RR} )_{11},\nonumber\\
&&C_3^{p, S_1}= (V^T y^{LL}_1)_{11} (z^{RR}_{1})_{11}, \nonumber\\
&& C_4^{p, S_1}=  (z^{RR}_1)_{11}  (y_1^{RR} )_{11}.
\label{Cptree}
\end{eqnarray}
Due to the neutrino in the final state, only $C_{1}^{n,S_{1}}$ and $C_{3}^{n,S_{1}}$  are nonzero
 \begin{eqnarray}
&&C_1^{n,S_{1}}=   (z^{RR}_1)_{11} (V^T  y_1^{LL} )_{11} .\nonumber\\
\label{Cntree}
\end{eqnarray}
These processes are illustrated in Fig. \ref{Fig-ann}.
 \begin{figure}[ht]
  \includegraphics[width=.7\linewidth]{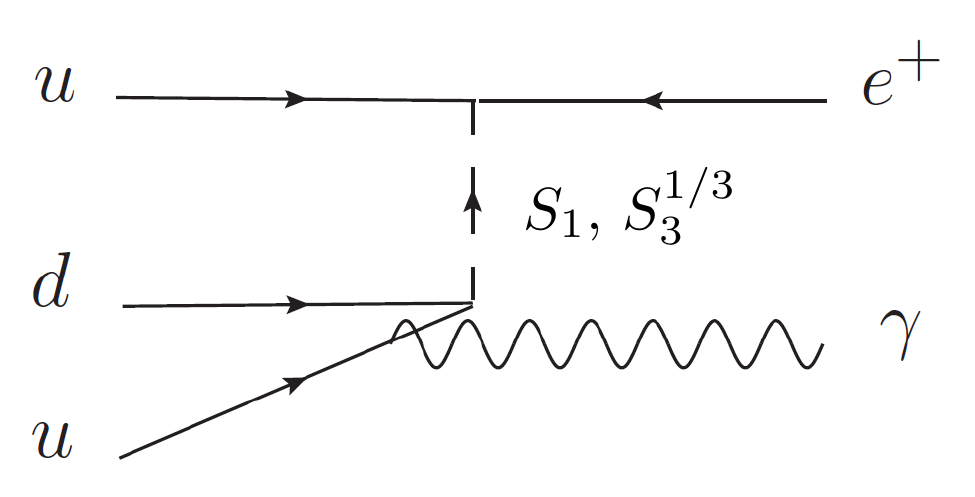}
\caption{$S_1$ mediating  proton decays  into $e^+$ and $\gamma$. The photon line should be attached to each charged particle. }
 \label{Fig-ann}
\end{figure}
The mixing parameters are then 
\begin{eqnarray}
&& \varepsilon^{p,S_1}_{1} = \frac{C^{p}_{1}  \beta_{p}}{m_{S_{1}}^{2}},\quad  \varepsilon^{p,S_1}_{2}=- \frac{C^{p}_{2}  \alpha_{p}}{m_{S_{1}}^{2}}, \quad  \varepsilon^{p,S_1}_{3}= \frac{C^{p}_{3}  \alpha_{p}}{m_{S_{1}}^{2}}, \nonumber \\
&&\varepsilon^{p,S_{1}}_{4} = - \frac{C^{p}_{4}  \beta_{p}}{m_{S_{1}}^{2}}, \quad \varepsilon_{1}^{n, S_{1}} =- \frac{C^{n}_{1}  \beta_{n}}{m_{S_{1}}^{2}}. 
\label{epn}
\end{eqnarray}
 When the $S_{3}^{1/3}$ leptoquark mediates these processes, only $C_{1}$  Wilson coefficient contribute with 
 \begin{eqnarray}
 && C_{1}^{p,S_{3}} = (V^T y_{3}^{LL})_{11} (V^T z_{3}^{LL})_{11}, \nonumber\\
 &&C_{1}^{n,S_{3}}  = (y_{3}^{LL} U)_{11} (V^T z_{3}^{LL})_{11} . 
 \label{WC-S3}
 \end{eqnarray}
  The mixing parameters become 
 \begin{eqnarray}
&&\varepsilon_{1}^{p,S_{3}} = \frac{C^{p,S_{3}}_{1} \beta_{p}}{m_{S_{3}}^{2}},\quad \varepsilon_{1}^{n,S_{3} }=- \frac{C^{n,S_{3}}_{1} \beta_{n}}{m_{S_{3}}^{2}}. 
 \label{mix-S3}
 \end{eqnarray}

\subsection{ Loop diagrams  in  $\Delta B =1$ transitions}

In Ref. \cite{Dorsner:2012nq} we considered 
loop induced proton (neutron) decay diagrams in  the case of $\tilde{S}_1=(\overline{\mathbf{3}},\mathbf{1},4/3)$, 
illu\-strated in Figs. 2 left (right). 
\begin{figure}[H]
  \includegraphics[width=1\linewidth]{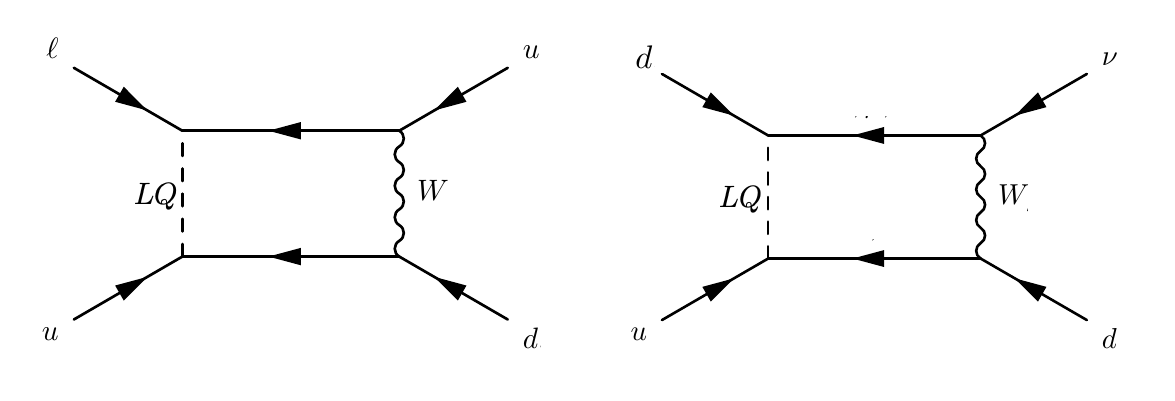}
 \label{pbox}
\caption{$p\to \ell^+ \pi^0$  (left) and $n \to \bar \nu \gamma$  (right) 
 mediated by one leptoquark, $LQ =  \tilde S_{1}$,  $S_{3}^{4/3}$, or $S_{3}^{-2/3}$, and $W$  in the box.} 
\end{figure}

The $\tilde S_1$ leptoquark has  couplings to a charged lepton and diquark coupling of the two up-type quarks from two different generations
\begin{eqnarray}
\mathcal{L}_{\tilde S_1}^{p} &=&-(\tilde y_1)_{ij} \bar{d}^{C\,i} \tilde S_{1} P_{R}\ell^{j}+(\tilde z_{1})_{{ij} } \bar{u}^{C\,i} \tilde S_{1}^{\ast} P_{R} u^{j} . \nonumber\\
&&+ \textrm{h.c.}
\label{L-tilde-S1}
\end{eqnarray}
The contribution of the diagram in Fig. 2 (left)  is
\begin{eqnarray}
&&C_{2}^{p,\tilde{S}_1}=  -\frac{G_{F}}{4 \pi^2 } \nonumber\\
&& \times \sum_{j,k} \tilde y_{1\,1k} \tilde z_{1\,1j} m_{u_{j}} m_{d_{k}}   V_{j1} V_{k1}^{\ast} 
x_{\tilde S_{1}} J(x_{\tilde S_{1}} , x_{u_{j}} , x_{d_{k}} ) 
\label{c-box-p}
\end{eqnarray}
In the case of a neutron $\Delta B=1$ decay, shown in Fig. 2 (right), the contribution is 
\begin{eqnarray}
&&C_{2}^{n,\tilde{S}_1}=-\frac{G_{F}}{4 \pi^2 } \sum_{j,i} \tilde y_{1\,1i} \tilde z_{1\,1j} m_{u_{j}} V_{j1}  m_{\ell_{i}} 
x_{\tilde S_{1}} J(x_{\tilde S_{1}} , x_{u_{j}} , x_{\ell_{i}} ), \nonumber\\
 \label{c-box-n}
\end{eqnarray}
with $x_{n}\equiv m_{n}^{2}/ m_{W}^{2}$ and $J(x,y,z)$. 
The mass dependence, apart from helicity
flip factors is encoded in the function $J(x,y,z)$ as 
\begin{eqnarray}
J(x,y,z) = &&\frac{(y -4) y  \log y}{(y -1)
     (y -x) (y-z)}+\frac{(z-4) z \log z}{(z-1) (z-y) (z-x
   )}\nonumber\\
&&+\frac{(x-4) x \log x}{(x-1) (x -y) (x-z)}.\nonumber\\
\end{eqnarray}
The mixing coefficients are $\varepsilon_{2} ^{p,\tilde S_{1}}= C_{2}^{p,\tilde{S}_1}\alpha_{p}/m_{\tilde S_{1}}^{2}$  and 
$\varepsilon_{2}^{n,\tilde S_{1}}= C_{2}^{n,\tilde{S}_1} \alpha_{n}/m_{\tilde S_{1}}^{2} $.

The triplet leptoquarks $S_{3}^{4/3}$ and $S_{3}^{-2/3}$ can generate the loop-induced nucleon decay as $\tilde S_{1}$. Due to the chirality difference of the operators in the amplitude,  we calculate the loop contribution using the $\xi =1$ gauge and find a negligible contribution of the ghost field. In the case of the $S_{3}^{-2/3}$ leptoquark, the diquark coupling is between different generations of the two down-type quarks. However, the amplitude is the same as in the case of $S_{3}^{4/3}$.

\begin{eqnarray}
&& C_1^{p,S_{3}}= \frac{1}{4\pi^2} \frac{G_{F}m_{S_3}^{2 }}{\sqrt{2} } \sum_{D=d,s,b} y_{3\,1D}^{LL}\sum_{U=c,t} ( V^T z_{3}^{LL} V^\dag)_{1U}  \nonumber\\
&&V_{uD}^{\ast}V_{Ud}
 \left[ \tilde J(x_U,x_D, x_{S_{3}}) +  \tilde J(x_D ,x_U ,  x_{S_{3}})  + \tilde J( x_{S_{3}} x_U,  x_D)  \right]. \nonumber\\
\label{S3-Cp}
\end{eqnarray}
We determine the box function 
$\tilde J(x,y,z)= x^{2} \log x /((x-1)(x-y)(x-z))$.  In  the case of neutron decay, the replacement is 
$\sum_{D=d,s,b} y_{3\,1D}^{LL}\sum_{U=c,t} ( V^T z_{3}^{LL} V^\dag)_{1U} V_{uD}^{\ast}V_{Ud}
\to \sum_{{\ell= e,\mu,\tau}} y_{3\,1\ell }^{LL}\sum_{U=c,t} ( V^T z_{3}^{LL} V^\dag)_{1U} V_{Ud}$.

\subsection{Triple leptoquarks coouplings}

Recently in Ref. \cite{Dorsner:2022twk}, we considered triple-leptoquark interactions for proton decay modes that arise at the tree- and one-loop levels.  Despite the usual loop-suppression factor, we found that the one-loop level decay amplitudes are much more relevant than the tree-level ones for the proton decay signatures. In this study, it is essential that diquark coupling with the leptoquark can be generated by the penguin operator, as presented in Fig. \ref{Figpeng} .
We consider three leptoquark mass eigenstates $S^Q$, $S^{Q^\prime}$ and $S^{Q^{\prime\prime}}$, where the superscripts denote the electric charges of each state, which satisfy $Q+Q^\prime+Q^{\prime\prime}=0$. \begin{align}
\label{eq:app-mix}
\mathcal{L}_\mathrm{scalar}= \lambda\, v\, \epsilon^{\alpha \beta \gamma} \, S_{\alpha}^Q\, S_\beta^{Q^\prime}\, S_\gamma^{Q^{\prime\prime}} +\mathrm{h.c.}\,,
\end{align}
The coupling $\lambda$ can be easily identified for each scenario as presented in Table~2.1  of Ref. \cite{Dorsner:2022twk}.
The fermion numbers  of the leptoquarks $S^Q$ and $S^{Q^\prime}$ are  $F=0$ and $F=2$, respectively with the general Yukawa interactions
\begin{eqnarray}
\label{eq:app-yuk}
&&\mathcal{L}_\mathrm{yuk.} = \overline{q} \left( y_{R} P_R + y_{L} P_L \right) \ell  \,S^{Q}\nonumber\\
&& + \overline{q^{\prime C}}
 \left( y_{R}^\prime P_R + y_{L}^\prime P_L \right) \ell  \,S^{Q^{\prime}\ast} +\mathrm{h.c.}\,,
\end{eqnarray}

\noindent in addition to the Yukawa couplings of $S^{Q^{\prime\prime}}$. In the above equations 
$\ell$ is a generic lepton, and $q$ and $q^\prime$ stand for two distinct quarks, with electric charges satisfying $Q=Q_q - Q_\ell$ and $Q^\prime=-Q_{q^\prime}-Q_\ell$. 
The colour and flavour indices are not explicitly written in Eq.~\eqref{eq:app-yuk}. 

The loop diagram corresponds to a loop-induced diquark coupling of the $S^{Q^{\prime\prime}}$ leptoquark,
\begin{align}
\label{eq:app-diquark}
\mathcal{L}_\mathrm{qq^\prime} = \epsilon^{\alpha \beta \gamma}\,\overline{q_\alpha^C} \left(y_{q q^\prime}^L P_L + y_{q q^\prime}^L P_R \right)q_\beta^\prime \, S_\gamma^{Q^{\prime\prime}} +\mathrm{h.c.}\,,
\end{align}

\noindent where $y_{q q^\prime}^L$ and $y_{q q^\prime}^R$ are explained in detail in Ref. \cite{Dorsner:2022twk} 
\begin{equation}
y_{q q^\prime}^L = \dfrac{\lambda v}{16 \pi^2 m_S^2} \left( m_\ell \, y_L^\prime y_R^\ast  \right) , \,
y_{q q^\prime}^R = \dfrac{\lambda v}{16 \pi^2 m_S^2} \left( m_\ell \, y_R^\prime y_L^\ast \right) .
\label{Triple-yuk}
\end{equation}
In these expressions, leptoquark masses are assumed to be degenerate, i.e., $m_{S^Q} =m_{S^{Q^\prime}} =m_{S^{Q^{\prime\prime}}} \equiv m_S $. The terms proportional to $m_\ell$  appear due to a chirality-flip in the internal lepton contribution. We do not write terms proportional to masses of $u, d$ quarks on the external legs in the case under consideration. Using these results, it is easy to determine the annihilation contribution of the nucleon to lepton. 
Table \ref{table2} presents contributions from the triple leptoquarks couplings. 

\begin{table}[tbp]
\centering
\begin{tabular}{|c|c|}
\hline 
Contractions & Process\\
\hline
$\tilde R_{2}$-$\tilde R_{2}$-$S_{1}^{\ast}$ & $n\to\bar \nu \gamma$\\
$R_{2}$-$\tilde R_{2}$- $\tilde S_{1}^{\ast}$ & $n\to \bar \nu \gamma$\\
$S_1$-$S_3$-$R_2^\ast$-$H$ & $p \to \ell^{+} \gamma$,\, $n\to \bar \nu \gamma$\\
$S_3$-$S_3$-$R_2^\ast$-$H$ & $p \to \ell^{+} \gamma$,\, $n\to \bar \nu \gamma$\\
$S_1$-$\tilde{S}_1$-$R_2^\ast$-$H^\ast$& $p \to \ell^{+} \gamma$,\, $n\to \bar \nu \gamma$\\
$S_3$-$\tilde{S}_1$-$R_2^\ast$-$H^\ast$& $p \to \ell^{+} \gamma$,\, $n\to \bar \nu \gamma$\\
$S_1$-$S_3$-$\tilde{R}_2^\ast$-$H^\ast$ & $p \to \ell^{+} \gamma$,\, $n\to \bar \nu \gamma$\\
$S_3$-$S_3$-$\tilde{R}_2^\ast$-$H^\ast$&  $p \to \ell^{+} \gamma$,\, $n\to \bar \nu \gamma$\\
\hline
\end{tabular} 
\caption{List of all non-trivial $LQ_{1}$-$LQ_{2}$-$LQ_{3}$-($H$) contractions and schematic representation of the associated $d=9$ effective operators generating the proton annihilation to a charged lepton.}
\label{table2}
\end{table}

As an example we consider  following interaction $ \lambda H^\dag i \tau_2 (\vec \tau \cdot \vec S_3 )^\ast  (\vec \tau \cdot \vec S_3 )^\ast i \tau_2 R_2 $ which leads after vacuum condensation to the Lagrangian 
\begin{eqnarray}
&& \mathcal L ( S_3 S_3 R_2^{\ast}) =\lambda v \sqrt{2} \epsilon^{\alpha \beta \gamma} \times\nonumber\\
&& \left(\sqrt{2} S_{3\alpha}^{-1/3} \, S_{3\beta}^{-4/3} \, R_{2\gamma}^{5/3} -  S_{3\alpha}^{-4/3} \, S_{3\beta}^{2/3} \, R_{2\gamma}^{2/3}\right) .
 \label{3LQ-0}
 \end{eqnarray}

\begin{figure}[htb!]
\begin{center}
  \centering
  \includegraphics[width=.7\linewidth]{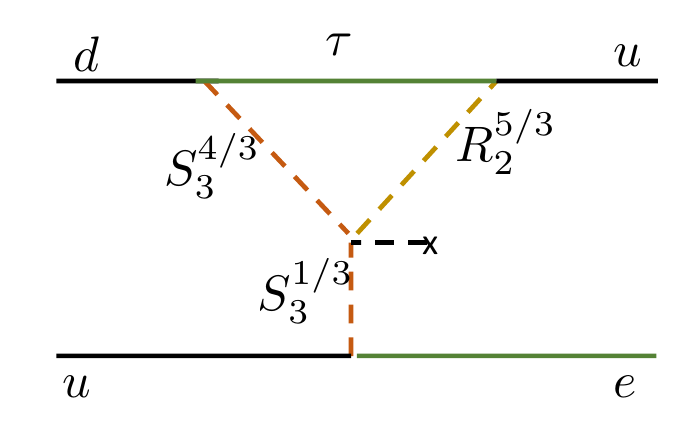}
  \end{center}
\caption{The annihilation diagram $u\, u\,d  \to e^+$, induced by the triple leptoquark couplings.}
 \label{Figpeng}
\end{figure}
Starting with the penguin-like diagram \cite{Dorsner:2022twk} we reduce the problem on the effective dimension-6  Lagrangian. 
 After integrating leptoquarks, we obtain the effective Lagrangian and assume that masses of $S_3$, $R_2$ equal $\Lambda$. 
At the scale $\Lambda =m_S$, for the $\tau$ lepton in the loop, when neglecting the contributions of the order $m_{u,d}/m_\tau$ the Wilson coefficient is
\begin{align}
C_{1}^{p,3LQ} &= \dfrac{\sqrt{2}\lambda }{8\pi^2 } \dfrac{v \, m_\tau}{m_S^2} (V^\ast y^L_{S_3}) y^L_{S_3} (V y_{R_2}^R)^\ast . 
\label{32}
\end{align}
Obviously, the mixing parameter in this case is $\varepsilon_{1}^{p,3LQ}= C_{1}^{p,3LQ} \beta_{p}/m_S^2$. 


 \section{ Numerical results }
\label{sec:4}

We use the recent lattice QCD results to calculate the decay widths of $p\to \ell^+ \pi^0$ to  $p\to \ell^+ \gamma$ ($\ell=e, \mu$). In Ref. \cite{Aoki:2017puj} the authors calculate 
 $\alpha_N = -0.0144(3)(21)$ GeV$^3$ and $\beta_N = 0.0144(3)(21)$ GeV$^3$, for $N=p,\, n$.  The form factors are $W_0^{RL}= 0.130$ GeV$^2$ for the matrix element \break $\langle \pi^0| (ud)_R u_L| p\rangle$  and $W_0^{LL}= 0.134(5)(16)$ GeV$^2$ for   \break $\langle \pi^0| (ud)_L u_L| p\rangle$ ($W_0^{LL}= W_0^{RR}$).

\begin{figure}[htb!]
\begin{center}
  \centering
  \includegraphics[width=0.8\linewidth]{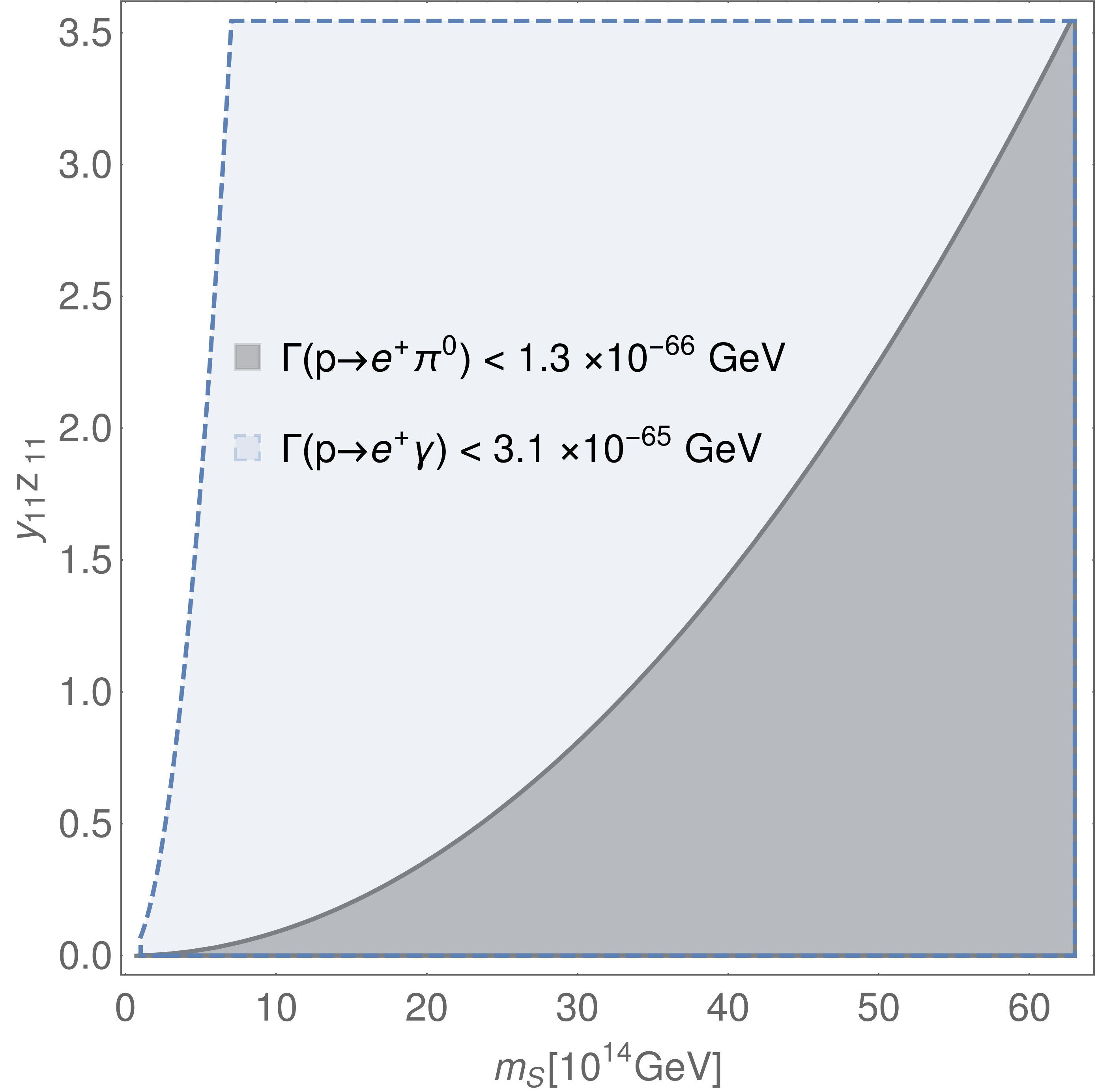}
  \end{center}
\caption{Allowed parameter space for the products of two Yukawa couplings and mass of a scalar leptoquark $S_1$ or $S_3$ for the tree-level $\Delta B =1$ transition. We use the experimental bounds on   $\Gamma(p\to e^+ \pi^0) $.  For comparison, we show much weaker bounds from $\Gamma(p\to e^+ \gamma )$.}
 \label{Tree}
\end{figure}


\begin{figure}[htb!]
\begin{center}
  \centering
  \includegraphics[width=0.8\linewidth]{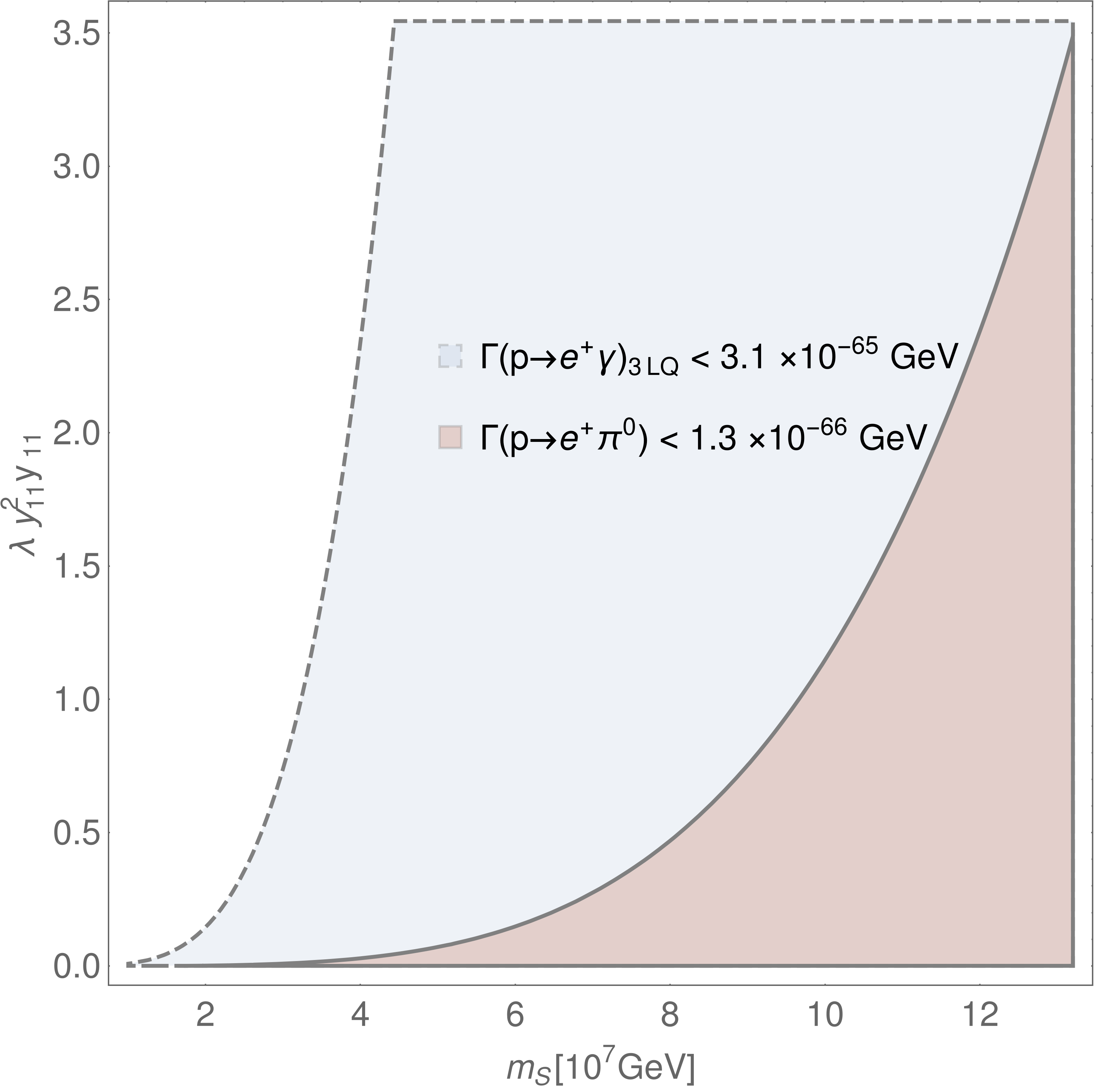}
  \end{center}
\caption{Allowed parameter space for the product of triple leptoquark coupling $\lambda$ with two Yukawa couplings and mass of a scalar leptoquark, assuming $m_{S_3^{-4/3}} = m_{S_3^{2/3}}= m_{R_2^{2/3}}$ coming from  $\Gamma(p\to e^+ \pi^0) $.  We show much weaker bounds from $\Gamma(p\to e^+ \gamma)$ for comparison. The label $\lambda\, y_{11}^{2} y_{11}$ stands for 
$ \lambda (V^\ast y^L_{S_3}) y^L_{S_3} (V y_{R_2}^R)^\ast$  in Eq. (\ref{32}).}
 \label{Triple}
\end{figure} 

In Fig. \ref{Tree}, we present constraints on the product of the Yukawas in Eq. (\ref{Cptree}) and the mass of a scalar leptoquark. We assume that the product of Yukawas is not larger than the perturbativity limit. As discussed in Section \ref{sec:2}, the bounds for Yukawas and the mass of leptoquarks are much weaker for the radiative mode. The flavour physics might constrain some of the product Yukawas as shown in \cite{Dorsner:2016wpm}. In flavour physics, for the ``pure leptoquark couplings'',  the masses of leptoquarks are  in the TeV region.  By allowing diquark couplings, or triple leptoquark couplings, the scale has to be several orders of magnitude larger, or a product of Yukawa couplings should be extremely small. However, all constraints depend strongly on the assumptions used for the texture of the Yukawa matrices. There are no general fits for all Yukawa matrices for the leptoquarks used in this study. All studies use some assumptions on the texture \cite{Becirevic:2022tsj,Becirevic:2016yqi,Crivellin:2022mff,Gherardi:2020qhc,Carvunis:2021dss}.  

The experimental bounds on the nucleon radiative decay widths are poor compared to the decay widths of $N \to \ell \pi$ modes. The same statement holds for the box diagrams with $\tilde S_{1}$, $S_{3}^{4/3}$, and  $S_{3}^{-2/3}$, but the mass of both leptoquarks is for a factor of $100$ or $10$ smaller than in the case of a tree-level transition. 
 

In the case of triple leptoquarks coupling, by constraining the product of the Yukawa couplings to be in the range $0.001 <\lambda (V^\ast y^L_{S_3}) y^L_{S_3} (V y_{R_2}^R)^\ast < \sqrt{4\pi}$  and setting masses of leptoquarks to be equal, we present in Fig. \ref{Triple} the allowed parameter regions for bounds from $p\to e^+ \pi^0$, and indicate much worse bound from $p\to e^+ \gamma $.

\begin{table}[h!]
\centering
\begin{tabular}{|c|c|c|}
\hline 
$\ell$&scalar LQ  & LQ mass [GeV]  \\
\hline
& tree  $S_{1,3}^{1/3}$ &$  1.9 \times 10^{15}$ \\
$e$  &box $\tilde S_1^{4/3}$ & $ 1.5 \times 10^{12}$  \\
&box $S_3^{4/3}$, $S_3^{-2/3}$ & $2.7 \times 10^{14}$ \\
&triple LQ  & $ 7.2 \times 10^{7}$  \\
\hline
 &tree  $S_{1,3}^{1/3}$ &$  1.4 \times 10^{15}$ \\
$\mu$&box  $\tilde S_1^{4/3}$ & $ 1.1 \times 10^{12}$ \\
&box $S_3^{4/3}$, $S_3^{-2/3}$ & $2.0 \times 10^{14}$  \\
&triple LQ & $ 3.7 \times 10^{7}$  \\
\hline
\end{tabular} 
\caption{The masses of mediating leptoquarks (LQ) at tree-level, box diagram and penguin-like diagram with the three leptoquark coupling, which are obtained by satisfying the experimental bound on  $\Gamma(p\to e^+ \pi^0) < 1.3 \times 10^{-65}$  GeV, assuming that the products of Yukawa couplings are set to 1.
}
\label{table33}
\end{table}
Table \ref{table33} presents the masses of the mediating leptoquark, calculated from the best experimental bound on $\Gamma(p\to e^{+} \pi^{0})$, for tree-level, box diagram and triple leptoquarks coupling, assuming that the product of all Yukawas is set to be $1$.  We repeat the procedure for muon replacing positron in the final state. The experimental bound on $\Gamma (p\to \mu ^{+} \pi^{0})$ is used too.

\section{Summary}
\label{sec:5}
We revisited radiative nucleon decays exploring the approach of Ref. \cite{Fornal:2018eol}.  This approach relies on the photon radiation from a hadron and charged lepton, benefiting from the knowledge of nucleons' anomalous magnetic moments. 
Then we describe an annihilation of a nucleon to a lepton within this framework and use lattice QCD results for the hadronic matrix elements.  We find that the radiative decay widths can be related to the decay widths of a nucleon decaying to a lepton and a pion, independent of the decay mechanism. These relations hold for any of the three types of transition, tree, box, and triple leptoquark transitions. Our results are 
 \begin{eqnarray}
\Gamma(p\to e^{+} \gamma)&\simeq & 3.8 \times 10^{-3}\, \Gamma(p\to e^{+} \pi^{0}), \nonumber\\
\Gamma(p\to \mu^{+} \gamma)& \simeq&4.6\times 10^{-3}\, \Gamma(p\to  \mu^{+} \pi^{0}) , \nonumber\\
\Gamma(n\to \bar \nu \,\,\gamma)&\simeq &3.8\times 10^{-3}\, \Gamma(n\to  \bar \nu \pi^{0}).
\label{Grad-pmu}
\end{eqnarray}
Then, we considered decay amplitudes of $ N \to \ell \pi^{0}$ and $N\to\ell \gamma$ mediated by scalar leptoquarks. The leptoquark interaction can occur on the tree- and loop levels. In the case of loop transitions, there are box-diagram and a transition via triple-leptoquarks interactions. The box diagram containing $\tilde S_{1}$ was known already. We completed the analysis of these transitions by calculating the box diagram contributions coming from the $S_{3}^{4/3}$ or $S_{3}^{-2/3}$ leptoquarks. We can predict the mass range of mediating scalar leptoquark by using the existing bound on the decay width of $p\to \ell^{+} \pi^{0}$ ($n\to \bar \nu \pi^{0}$).  In the case of tree-level transition, by assuming the products of the leptoquark Yukawa couplings to be of order 1, the mass of $S_{1,3}^{1/3}$ reaches $10^{15}$ GeV. In comparison, the box transition can reduce it to $10^{12}-10^{14}$ GeV, depending on the baryon number violating operator.  In the case of a triple leptoquark interaction destabilising nucleon, the mass scale of the leptoquark is further reduced to the order of $10^{7}$ GeV.
The existing experimental bounds for $\Gamma(N\to\ell \gamma)$ are suppressed by a factor of 20 compared to the experimental bounds for $\Gamma(N \to \ell \pi)$. Our calculations give $10^{3}$ factor suppression compared to the widths of $N \to \ell \pi$. This can be useful guidance for further experimental studies of nucleon radiative decays.

\noindent
{\bf  Acknowledgements}
S. F. acknowledges the financial support from the Slovenian Research Agency (research core funding No. P1-0035). 
M. S. is supported by Slovenian Research Agency ARIS (Grant No. 53647).
\bibliographystyle{elsarticle-num}
\bibliography{current-p}
\end{document}